\documentclass[10pt,conference]{IEEEtran}

\usepackage[cmex10]{amsmath}
\usepackage{amssymb}
\usepackage{bm}

\usepackage{algpseudocode}
\usepackage{algorithm}

\usepackage{array}

\usepackage{url}

\usepackage[noadjust]{cite}

\newtheorem{proposition}{Proposition}

\usepackage{fancyhdr}
\usepackage{amsfonts}
\usepackage{xfrac}
\usepackage{color}
\usepackage{pgfplots}
\pgfplotsset{
  compat=1.17,
  plot coordinates/math parser=false,
  filter discard warning=false,
  table/col sep=space
}
\usepackage{enumerate}
\usepackage{morewrites}
\usepackage{morefloats}
\usepackage{expl3}
\usepackage{shellesc}
\usepackage{mathrsfs}
\usepackage[font=footnotesize,labelfont=bf]{caption}
\usepackage{subcaption}
\usepackage{stfloats}
\usepackage{makecell}
\usepackage{comment}
\usepackage{afterpage}
\usepackage{tikz}
\usepackage{mathdots}
\usepackage{yhmath}
\usepackage{cancel}
\usepackage{siunitx}
\usepackage{array}
\usepackage{multirow}
\usepackage{textcomp}
\usepackage{gensymb}
\usepackage{tabularx}
\usepackage{booktabs}
\usepackage{graphicx}
\usepackage{xcolor}
\usepackage{mathtools}

\allowdisplaybreaks
\usetikzlibrary{decorations.markings,decorations.pathmorphing}

\begin{document}

\title{Ergodic Rate Analysis of Two-State \\ Pinching-Antenna Systems}

\author{\IEEEauthorblockN{Dimitrios Tyrovolas\IEEEauthorrefmark{1}\IEEEauthorrefmark{4},
                         Sotiris Tegos\IEEEauthorrefmark{2}, Yue Xiao\IEEEauthorrefmark{3},
                         Panagiotis Diamantoulakis\IEEEauthorrefmark{2},\\
                         Sotiris Ioannidis\IEEEauthorrefmark{4}\IEEEauthorrefmark{5},
                         Christos Liaskos\IEEEauthorrefmark{6},
                         George Karagiannidis\IEEEauthorrefmark{2}, and Stylianos D. Asimonis\IEEEauthorrefmark{1}}
\IEEEauthorblockA{\IEEEauthorrefmark{1}Department of Electrical and Computer Engineering,
                  University of Patras,
                  26504 Patras, Greece\\
                  Email: \{dtyrovolas, s.asimonis\}@upatras.gr}
\IEEEauthorblockA{\IEEEauthorrefmark{2}Department of Electrical and Computer Engineering,
                  Aristotle University of Thessaloniki,
                  54124 Thessaloniki, Greece\\
                  Email: \{tegosoti, padiaman, geokarag\}@auth.gr}
\IEEEauthorblockA{\IEEEauthorrefmark{3}School of Information Science and Technology,
                  Southwest Jiaotong University, Chengdu 610032\\
                  Email: xiaoyue@swjtu.edu.cn}
\IEEEauthorblockA{\IEEEauthorrefmark{4}Dienekes SI IKE, 71414 Heraklion, Greece}
\IEEEauthorblockA{\IEEEauthorrefmark{5}Department of Electrical and Computer Engineering,
                  Technical University of Chania,
                  73100 Chania, Greece\\
                  Email: sotiris@ece.tuc.gr}
\IEEEauthorblockA{\IEEEauthorrefmark{6}Computer Science Engineering Department,
                  University of Ioannina,
                  45110 Ioannina, Greece\\
                  Email: cliaskos@uoi.gr}
}


\maketitle	

\begin{abstract} 
Flexible Antenna Systems (FAS) are a key enabler of next-generation wireless networks, allowing the antenna aperture to be dynamically reconfigured to adapt to channel conditions and service requirements. In this context, pinching-antenna systems (PASs) implemented on software-controllable dielectric waveguides provide the ability to reconfigure both channel characteristics and path loss by selectively exciting discrete radiation points. Existing works, however, typically assume continuously adjustable pinching positions, neglecting the spatial discreteness imposed by practical implementations. This paper develops a closed-form analytical framework for the ergodic rate of two-state PASs, where pinching antennas are fixed and only their activation states are controlled. To quantify the impact of spatial discretization, pinching discretization efficiency is introduced, characterizing the performance gap relative to the ideal continuous case. Finally, numerical results show that near-continuous performance can be achieved with a limited number of pinching points, providing design insights for scalable PASs.
\end{abstract}

\maketitle

\section{Introduction}

The evolution of next-generation wireless communication networks involves transitioning from static antenna configurations to flexible and adaptive antenna architectures capable of adapting in real time to varying service requirements \cite{6GNetwork}. In this emerging framework, the concept of Flexible Antenna Systems (FAS) has introduced a new communication paradigm in which the radiating structure becomes a reconfigurable physical-layer resource rather than a fixed hardware component \cite{FAS2021,Ding2024TCOM}. By embedding software-controlled reconfigurability directly into the antenna aperture, FAS enables dynamic adjustment of signal characteristics, such as directionality, attenuation, and coverage under software-defined control \cite{FAS2021,Assimonis}. Among the technologies envisioned to realize FAS, pinching antennas (PAs) have recently emerged as a distinctive approach capable of reconfiguring both small-scale channel variations and the large-scale path loss of wireless links \cite{Ding2024TCOM,DOCOMO}. In particular, by locally exciting radiating points along a software-controllable dielectric waveguide, PAs can flexibly adapt the effective propagation distance and radiation footprint, thus introducing the ability to program the antenna–channel interaction according to user location and environmental geometry \cite{TegosPinching}. Therefore, it becomes essential to examine the performance capabilities and inherent limitations of dielectric-waveguide-based flexible antenna systems under practical deployment conditions to establish a realistic understanding of their achievable benefits.

Building upon this emerging concept, a growing body of research has investigated the fundamental characteristics and potential of pinching-antenna systems (PASs) as a specific realization of flexible antenna systems across different scenarios. Initially, several works have established analytical models describing the behavior of electromagnetic radiation along the reconfigurable dielectric waveguide, leading to optimized PA-based beamforming schemes and revealing significant improvements in performance \cite{PASS, Modeling2025}. Additionally, the authors in \cite{TyrovolasPASS2025} and \cite{VasilisPASS} have characterized key performance metrics, such as outage probability, rate, and power transfer efficiency, offering insight into how the waveguide length and dielectric losses of the flexible antenna structure shape system behavior. Finally, recent studies have explored the potential of PASs in integrated sensing and communication \cite{BozanisPASS}, physical layer security \cite{PIGIPASS}, and multi-user communication scenarios \cite{ThrassosPASS,Kaikit2025,APOSTOLOSRSMA}, confirming the versatility of PASs as a strong candidate for flexible antenna services in 6G networks. Despite these advances, most existing works rely on the idealized assumption that PAs can be continuously adjusted along the waveguide, while in practice only a finite number of PAs can be formed on a discretely controllable dielectric waveguide, which complicates the analytical derivation of performance metrics. To the best of the authors’ knowledge, no work has yet provided a closed-form mathematical framework that explicitly quantifies the impact of such spatial discretization on the performance of PASs.

In this direction, in this work we analyze the performance of two-state PASs, where the positions of the PAs are fixed along the dielectric waveguide and only their activation state can be controlled. By incorporating the spatial discreteness of the available pinching points, we derive a closed-form analytical expression for the ergodic data rate, providing an exact characterization of the achievable data rate performance. Furthermore, we introduce pinching discretization efficiency (PDE) that quantifies the performance gap between the discrete and continuous pinching configurations, enabling a direct assessment of the number of PAs required to approximate the ideal continuous case. As a result, this work establishes a mathematical framework connecting the number of available PAs, the system geometry, and the resulting rate performance, offering valuable insights for the efficient design of PASs within PWEs.

\section{System model}

\begin{figure}
 \centering
    \resizebox{\columnwidth}{!}{
\begin{tikzpicture}[x=0.75pt,y=0.75pt,yscale=-1,xscale=1]
\draw   (238.67,50) -- (566,50) -- (425.71,226.6) -- (98.38,226.6) -- cycle ;
\draw  [dash pattern={on 4.5pt off 4.5pt}]  (98.38,226.6) -- (266.13,16.35) ;
\draw [shift={(268,14)}, rotate = 128.58] [fill={rgb, 255:red, 0; green, 0; blue, 0 }  ][line width=0.08]  [draw opacity=0] (8.93,-4.29) -- (0,0) -- (8.93,4.29) -- cycle    ;
\draw  [fill={rgb, 255:red, 235; green, 232; blue, 233 }  ,fill opacity=1 ] (530,88) -- (197.97,88.87) -- (198,98) -- (530.03,97.13) -- cycle ;
\draw  [fill={rgb, 255:red, 248; green, 231; blue, 28 }  ,fill opacity=1 ] (351,85) -- (366.5,85) -- (366.5,100.5) -- (351,100.5) -- cycle ;
\draw (317.75,176) node  {\includegraphics[width=28.13pt,height=22.5pt]{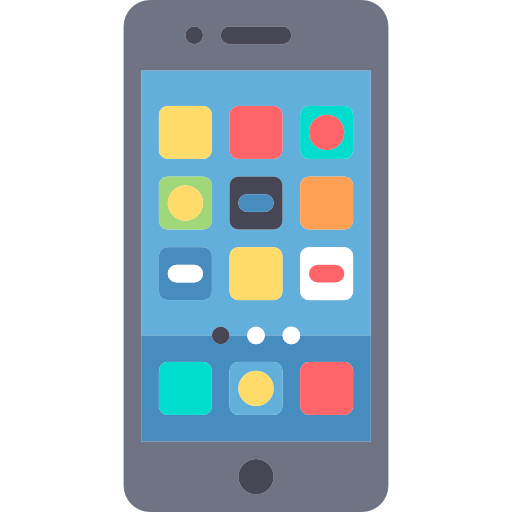}};
\draw  [dash pattern={on 4.5pt off 4.5pt}]  (172,134.5) -- (411.78,135.07) -- (498,134.52) ;
\draw [shift={(501,134.5)}, rotate = 179.63] [fill={rgb, 255:red, 0; green, 0; blue, 0 }  ][line width=0.08]  [draw opacity=0] (8.93,-4.29) -- (0,0) -- (8.93,4.29) -- cycle    ;
\draw    (215,105) -- (196.5,105) ;
\draw [shift={(194.5,105)}, rotate = 360] [color={rgb, 255:red, 0; green, 0; blue, 0 }  ][line width=0.75]    (10.93,-3.29) .. controls (6.95,-1.4) and (3.31,-0.3) .. (0,0) .. controls (3.31,0.3) and (6.95,1.4) .. (10.93,3.29)   ;
\draw    (214,105) -- (238,105) ;
\draw [shift={(240,105)}, rotate = 180] [color={rgb, 255:red, 0; green, 0; blue, 0 }  ][line width=0.75]    (10.93,-3.29) .. controls (6.95,-1.4) and (3.31,-0.3) .. (0,0) .. controls (3.31,0.3) and (6.95,1.4) .. (10.93,3.29)   ;
\draw  [draw opacity=0] (145.25,147) .. controls (145.25,144.65) and (147.15,142.75) .. (149.5,142.75) .. controls (151.85,142.75) and (153.75,144.65) .. (153.75,147) .. controls (153.75,149.35) and (151.85,151.25) .. (149.5,151.25) .. controls (147.15,151.25) and (145.25,149.35) .. (145.25,147) -- cycle ;
\draw (179.5,90.75) node  {\includegraphics[width=66.75pt,height=47.63pt]{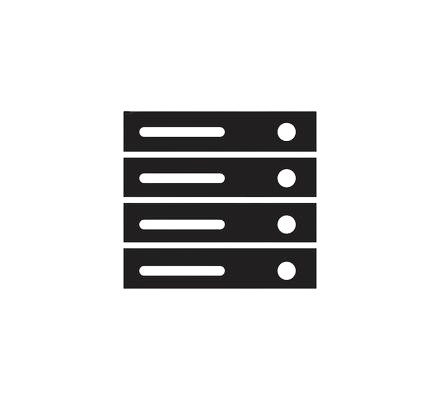}};
\draw  [fill={rgb, 255:red, 0; green, 0; blue, 0 }  ,fill opacity=1 ] (241,85) -- (256.5,85) -- (256.5,100.5) -- (241,100.5) -- cycle ;
\draw    (278,105) -- (259.5,105) ;
\draw [shift={(257.5,105)}, rotate = 360] [color={rgb, 255:red, 0; green, 0; blue, 0 }  ][line width=0.75]    (10.93,-3.29) .. controls (6.95,-1.4) and (3.31,-0.3) .. (0,0) .. controls (3.31,0.3) and (6.95,1.4) .. (10.93,3.29)   ;
\draw    (277,105) -- (301,105) ;
\draw [shift={(303,105)}, rotate = 180] [color={rgb, 255:red, 0; green, 0; blue, 0 }  ][line width=0.75]    (10.93,-3.29) .. controls (6.95,-1.4) and (3.31,-0.3) .. (0,0) .. controls (3.31,0.3) and (6.95,1.4) .. (10.93,3.29)   ;
\draw    (325,105) -- (306.5,105) ;
\draw [shift={(304.5,105)}, rotate = 360] [color={rgb, 255:red, 0; green, 0; blue, 0 }  ][line width=0.75]    (10.93,-3.29) .. controls (6.95,-1.4) and (3.31,-0.3) .. (0,0) .. controls (3.31,0.3) and (6.95,1.4) .. (10.93,3.29)   ;
\draw    (325,105) -- (349,105) ;
\draw [shift={(351,105)}, rotate = 180] [color={rgb, 255:red, 0; green, 0; blue, 0 }  ][line width=0.75]    (10.93,-3.29) .. controls (6.95,-1.4) and (3.31,-0.3) .. (0,0) .. controls (3.31,0.3) and (6.95,1.4) .. (10.93,3.29)   ;
\draw    (387,105) -- (368.5,105) ;
\draw [shift={(366.5,105)}, rotate = 360] [color={rgb, 255:red, 0; green, 0; blue, 0 }  ][line width=0.75]    (10.93,-3.29) .. controls (6.95,-1.4) and (3.31,-0.3) .. (0,0) .. controls (3.31,0.3) and (6.95,1.4) .. (10.93,3.29)   ;
\draw    (387,105) -- (411,105) ;
\draw [shift={(413,105)}, rotate = 180] [color={rgb, 255:red, 0; green, 0; blue, 0 }  ][line width=0.75]    (10.93,-3.29) .. controls (6.95,-1.4) and (3.31,-0.3) .. (0,0) .. controls (3.31,0.3) and (6.95,1.4) .. (10.93,3.29)   ;
\draw    (434,105) -- (415.5,105) ;
\draw [shift={(413.5,105)}, rotate = 360] [color={rgb, 255:red, 0; green, 0; blue, 0 }  ][line width=0.75]    (10.93,-3.29) .. controls (6.95,-1.4) and (3.31,-0.3) .. (0,0) .. controls (3.31,0.3) and (6.95,1.4) .. (10.93,3.29)   ;
\draw    (434,105) -- (458,105) ;
\draw [shift={(460,105)}, rotate = 180] [color={rgb, 255:red, 0; green, 0; blue, 0 }  ][line width=0.75]    (10.93,-3.29) .. controls (6.95,-1.4) and (3.31,-0.3) .. (0,0) .. controls (3.31,0.3) and (6.95,1.4) .. (10.93,3.29)   ;
\draw  [fill={rgb, 255:red, 0; green, 0; blue, 0 }  ,fill opacity=1 ] (459,85) -- (474.5,85) -- (474.5,100.5) -- (459,100.5) -- cycle ;
\draw    (495,105) -- (476.5,105) ;
\draw [shift={(474.5,105)}, rotate = 360] [color={rgb, 255:red, 0; green, 0; blue, 0 }  ][line width=0.75]    (10.93,-3.29) .. controls (6.95,-1.4) and (3.31,-0.3) .. (0,0) .. controls (3.31,0.3) and (6.95,1.4) .. (10.93,3.29)   ;
\draw    (495,105) -- (519,105) ;
\draw [shift={(521,105)}, rotate = 180] [color={rgb, 255:red, 0; green, 0; blue, 0 }  ][line width=0.75]    (10.93,-3.29) .. controls (6.95,-1.4) and (3.31,-0.3) .. (0,0) .. controls (3.31,0.3) and (6.95,1.4) .. (10.93,3.29)   ;
\draw  [dash pattern={on 4.5pt off 4.5pt}]  (348,49.5) -- (209.38,224.6) ;
\draw  [dash pattern={on 4.5pt off 4.5pt}]  (456,50.5) -- (317.38,225.6) ;

\draw (111,124.99) node [anchor=north west][inner sep=0.75pt]    {$( 0,0,0)$};
\draw (476.2,132.17) node [anchor=north west][inner sep=0.75pt]    {$x$};
\draw (242,6.72) node [anchor=north west][inner sep=0.75pt]    {$y$};
\draw (352.4,63.89) node [anchor=north west][inner sep=0.75pt]   [align=left] {PA};
\draw (229.7,195.89) node [anchor=north west][inner sep=0.75pt]  [font=\footnotesize]  {$\boldsymbol{\psi }_{m} =\left( x_{m} ,y_{m} ,0\right)$};
\draw (167.5,57) node [anchor=north west][inner sep=0.75pt]   [align=left] {AP};
\draw (98,83.99) node [anchor=north west][inner sep=0.75pt]    {$( 0,0,h)$};
\draw (199.75,108.4) node [anchor=north west][inner sep=0.75pt]    {$\delta /2$};
\draw (257.75,109.4) node [anchor=north west][inner sep=0.75pt]    {$\delta /2$};
\draw (310.75,108.4) node [anchor=north west][inner sep=0.75pt]    {$\delta /2$};
\draw (367.75,108.4) node [anchor=north west][inner sep=0.75pt]    {$\delta /2$};
\draw (420.75,109.4) node [anchor=north west][inner sep=0.75pt]    {$\delta /2$};
\draw (475.75,109.4) node [anchor=north west][inner sep=0.75pt]    {$\delta /2$};
\end{tikzpicture}
}
    \caption{Overview of two-state PAS.}
    \label{fig:system_model}
\end{figure}
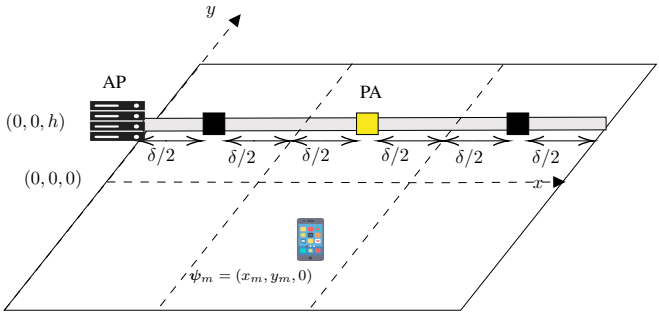

We consider the downlink communication scenario depicted in Fig.~\ref{fig:system_model}, where an access point (AP) communicates with a single-antenna user located randomly within a rectangular area in the $x$-$y$ plane with dimensions $D_x$ and $D_y$. The user position is denoted by $\boldsymbol{\psi_m} = (x_m, y_m, 0)$, where $x_m$ is uniformly distributed over $[0, D_x]$ and $y_m$ is uniformly distributed over $\left[-\frac{D_y}{2}, \frac{D_y}{2}\right]$. To ensure reliable communication, the AP employs a dielectric waveguide which allows electromagnetic radiation from selected points along its length through a controlled ``pinching'' mechanism. In more detail, the waveguide is oriented parallel to the $x$-axis at a height $h$, spanning a total length equal to $D_x$, and is equipped with $M$ PAs positioned at predefined locations where only one PA is activated during each transmission interval. In addition, the inter-spacing between consecutive PAs is denoted as $\delta = \frac{D_x}{M}$, and the coordinates of the $k$-th PA are given by $\boldsymbol{\psi_p^{(k)}} = (x_k, 0, h)$, with $x_k = \frac{2k-1}{2}\,\delta$ and $k = 1, 2, \ldots, M$. Therefore, the wireless channel between the $k$-th PA and the user is modeled as 
{\small
\begin{equation}
h_1^{(k)} = \frac{\sqrt{\eta} e^{-j \frac{2\pi}{\lambda} |\boldsymbol{\psi_m} -\boldsymbol{\psi_p^{(k)}}|}}{|\boldsymbol{\psi_m} -\boldsymbol{\psi_p^{(k)}}|},
\end{equation}
}where $\eta = \frac{\lambda^2}{16 \pi^2}$ denotes the path loss at a reference distance of 1 m, $\lambda$ is the free-space wavelength, $j$ is the imaginary unit, and $|\cdot|$ denotes the Euclidean norm. Moreover, as the signal propagates along the dielectric waveguide, it undergoes a phase shift determined by the effective refractive index $n_{\mathrm{eff}}$, which defines the guided wavelength as $\lambda_g = \frac{\lambda}{n_{\mathrm{eff}}}$. Accordingly, the phase shift accumulated from the waveguide feed point at $\boldsymbol{\psi_0} = (0, 0, h)$ to the $k$-th PA position is expressed as $h_2^{(k)} = e^{-j \frac{2\pi}{\lambda_g} |\boldsymbol{\psi_p^{(k)}} -\boldsymbol{\psi_0}|}$, where $\boldsymbol{\psi_0} = (0, 0, h)$ denotes the location of the waveguide feeding point. Consequently, the received signal at the user when the $k$-th PA is used can be expressed as 
\begin{equation}
\small
y_r = \sqrt{P_t}h_1^{(k)} h_2^{(k)} s + w_n,
\end{equation}
where $P_t$ is the transmit power, $s$ is the transmitted symbol with $\mathbb{E}[|s|^2] = 1$, and $w_n$ is additive white Gaussian noise with zero mean and variance $\sigma^2$. Therefore, the received SNR corresponding to the $k$-th PA is written as
{\small
\begin{equation}\label{SNR1_discrete}
\gamma^{(k)} = \frac{\eta P_t \left| e^{-j \left( \frac{2\pi}{\lambda} |\boldsymbol{\psi_m} -\boldsymbol{\psi_p^{(k)}}| + \frac{2\pi}{\lambda_g} |\boldsymbol{\psi_p^{(k)}} -\boldsymbol{\psi_0}| \right)} \right|^2}{\sigma^2 |\boldsymbol{\psi_m} -\boldsymbol{\psi_p^{(k)}}|^2}.
\end{equation}
}
Finally, considering that $|e^{-jx}| = 1$, \eqref{SNR1_discrete} simplifies to
{\small
\begin{equation}\label{SNR2_discrete}
\gamma^{(k)} = \frac{\eta P_t}{\sigma^2 |\boldsymbol{\psi_m} - \boldsymbol{\psi_p^{(k)}}|^2}
= \frac{\eta P_t}{\sigma^2 \big( (x_m - x_k)^2 + y_m^2 + h^2 \big)}.
\end{equation}
}

Since all PAs are distributed along the same waveguide, the access point activates the PA providing the maximum received SNR. Thus, considering that $\gamma^{(k)}$ in \eqref{SNR2_discrete} is a monotonically decreasing function of the distance between the user and the active PA, the optimal PA corresponds to the one that is closest to the user in the $x$-dimension. Therefore, by taking into account that $x_m$ is uniformly distributed over $[0, D_x]$, the horizontal distance between the user and the selected PA is defined as $\varepsilon = x_m - x_p$, where $\varepsilon$ follows a uniform distribution $\mathcal{U}\!\left[-\frac{\delta}{2}, \frac{\delta}{2}\right]$. Consequently, \eqref{SNR2_discrete} can be equivalently expressed as
{\small
\begin{equation}\label{SNR_eps}
\gamma_r = \frac{\eta P_t}{\sigma^2 \left( \varepsilon^2 + y_m^2 + h^2 \right)}.
\end{equation}
}
\section{Rate Analysis}
To quantify the performance of the considered two-state PAS, it is essential to determine its ergodic rate, which reflects the ergodic data throughput over the spatial distribution of the user. However, due to the discreteness of the available PA positions, the ergodic data rate exhibits a distinct spatial dependence that directly links the system geometry with its communication efficiency. In this direction, the following proposition provides a closed-form expression for the ergodic data rate, offering insights into how the inter-spacing $\delta$, the number of PAs, and the room dimensions jointly influence the rate performance.
\begin{proposition}
Considering that $\varepsilon \sim \mathcal{U}\!\left[-\frac{\delta}{2},\frac{\delta}{2}\right]$ and $y_m \sim \mathcal{U}\!\left[-\frac{D_y}{2},\frac{D_y}{2}\right]$, the ergodic rate of the considered two-state PAS can be expressed as
{\small
\begin{equation}\label{R_avg_f}
\overline{R}
= \frac{4}{\delta D_y \ln 2}\,\big(I_i(C+h^2) + I_j(C+h^2) - I_i(h^2) -I_j(h^2)\big),
\end{equation}}where $I_i(\cdot)$ and $I_j(\cdot)$ are given in \eqref{Iix} and \eqref{Ijx} at the top of the next page, respectively, $C=\frac{\eta P_t}{\sigma^2}$, $\mathrm{Ti}_2(z)=\tfrac{\mathrm{Li}_2(i z)-\mathrm{Li}_2(-i z)}{2i}$ denotes the arctangent integral function, with $\mathrm{Li}_2(\cdot)$ representing the dilogarithm function.
\end{proposition}

\begin{figure*}[t]
\begin{equation}\label{Iix}
\small
I_i(x)
= \frac{\delta D_y}{4}\,
    \ln\!\left(\frac{D_y^2}{4}+\frac{\delta^2}{4}+x\right)
   -\frac{\delta D_y}{2}
   +\delta\,\sqrt{\frac{\delta^2}{4}+x}\;
     \tan^{-1}\!\left(
       \frac{D_y}{2\sqrt{x+\frac{\delta^2}{4}}}
     \right).
\end{equation}
\hrule
\begin{equation}\label{Ijx}
\small
\begin{split}
I_j(x)
&=\Bigg[
     \frac{D_y}{2}\sqrt{x+\frac{D_y^2}{4}}
     + x
       \ln\!\left(\frac{D_y}{2}+\sqrt{x+\frac{D_y^2}{4}}\right)
  \Bigg]
  \tan^{-1}\!\left(
     \frac{\delta}{2\sqrt{x+\frac{D_y^2}{4}}}
  \right) -x\ln\!\big(\sqrt{x}\big)
     \tan^{-1}\!\left(\frac{\delta}{2\sqrt{x}}\right)+\frac{\delta D_y}{4}
  \\
&\quad
  -\frac{\delta}{2}\sqrt{x+\frac{\delta^2}{4}}\;
     \tan^{-1}\!\left(
        \frac{D_y}{2\sqrt{x+\frac{\delta^2}{4}}}
     \right) +x\Bigg[
     \Big(
        \ln\!\big(\sqrt{x}\big)
        + \operatorname{asinh}\!\Big(\frac{D_y}{2\sqrt{x}}\Big)
     \Big)
     \tan^{-1}\!\left(\frac{\sqrt{4x+D_y^2}}{\delta}\right)\\
&\qquad
      - \ln\!\big(\sqrt{x}\big)
       \tan^{-1}\!\Big(\frac{2\sqrt{x}}{\delta}\Big)
     - \frac{\pi}{2}\,
       \operatorname{asinh}\!\Big(\frac{D_y}{2\sqrt{x}}\Big)
     -\,\mathrm{Ti}_2\Big(
         e^{\operatorname{asinh}\!\left(\frac{D_y}{2\sqrt{x}}\right)}
         \frac{\delta}{2\sqrt{x}}\,
         \big(\sqrt{1+\tfrac{4x}{\delta^2}}-1\big)
       \Big) 
    \\
&\qquad
       +\,\mathrm{Ti}_2\Big(
         \frac{\delta}{2\sqrt{x}}\,
         \big(\sqrt{1+\tfrac{4x}{\delta^2}}-1\big)
       \Big) -\,\mathrm{Ti}_2\Big(
         -e^{\operatorname{asinh}\!\left(\frac{D_y}{2\sqrt{x}}\right)}
         \frac{\delta}{2\sqrt{x}}\,
         \big(\sqrt{1+\tfrac{4x}{\delta^2}}+1\big)
       \Big) +\,\mathrm{Ti}_2\Big(
         -\frac{\delta}{2\sqrt{x}}\,
         \big(\sqrt{1+\tfrac{4x}{\delta^2}}+1\big)
       \Big)
        \Bigg].
\end{split}
\end{equation}
\hrule
\end{figure*}

\begin{IEEEproof} 
Taking into account \eqref{SNR_eps}, the ergodic rate of the considered PAS can be expressed as
\begin{equation}\label{c1}
\overline{R}=\mathbb{E}\!\left[\log_2\!\left(1+\frac{C}{h^2+\varepsilon^2+y_m^2}\right)\right]
\end{equation}
with $\mathbb{E}[\cdot]$ denoting expectation. Since \eqref{c1} is an even function with respect to $\varepsilon$ and $y_m$, it can be written in integral form as follows
\begin{equation}
    \overline{R}=\frac{4}{\delta D_y\ln2} \int_0^{\frac{D_y}{2}} \int_0^{\frac{\delta}{2}} \ln \left(1+ \frac{C}{\varepsilon^2 + y_m^2 + h^2 } \right) d\varepsilon \, dy_m,
\end{equation}
which, after some algebraic manipulations, can be rewritten as
{\small
\begin{equation}\label{R_avg}
\overline{R}
= \frac{4}{\delta D_y \ln 2}\,\left(I_i(C+h^2) + I_j(C+h^2) - I_i(h^2) - I_j(h^2)\right),
\end{equation}}where $I_i(\cdot)$, and $I_j(\cdot)$ are equal to
\begin{equation}
    I_i(x) = \int_0^{\frac{D_y}{2}} \frac{\delta}{2} \ln\left( \frac{\delta^2}{4} + x +y^2_m\right) dy_m,
\end{equation}
and
\begin{equation}\label{Ij_13}
    I_j(x) = \int_0^{\frac{D_y}{2}} 2 \sqrt{x+ y^2_m} \tan^{-1} \! \left( \frac{\delta}{2 \sqrt{x+ y^2_m}}\right)dy_m.
\end{equation}
Thus, by following similar steps as shown in Appendix \ref{App:A}, we obtain \eqref{Iix} and \eqref{Ijx}, and by substituting them in \eqref{R_avg}, \eqref{R_avg_f} is derived, which concludes the proof.
\end{IEEEproof}

Based on the derived closed-form expression of the ergodic rate, we further define the PDE as
\begin{equation}\label{eq:ratio}
\small
\overline{\eta}_r=\frac{\overline{R}}{R_{\mathrm{c}}},
\end{equation}
where $R_{\mathrm{c}}$ corresponds to the ergodic data rate achieved by a single PA on an ideal continuous PAS in which a PA can be formed at any arbitrary point along the waveguide, whose expression is provided in \cite{Ding2024TCOM}. In this way, the PDE $\overline{\eta}_r$ indicates the relative performance loss introduced by the discrete PA configuration compared to its continuous PAS counterpart, completing the analytical rate characterization of the proposed two-state PAS.
\section{Numerical Results}
In this section, we evaluate the performance of the examined two-state PAS and the accuracy and validity of the derived expression. For consistency, the system parameters are chosen as in \cite{Ding2024TCOM}, where the noise power $\sigma^2$ is $-90$ dBm, the carrier frequency $f_c=28$ GHz, and the effective refractive index $n_{\mathrm{eff}}=1.4$. Moreover, the waveguide height is $h = 3$~m, and the deployment area is assumed to have dimension $D_y = 10$ m. Finally, to validate the theoretical results, Monte Carlo simulations are performed using $10^6$ random realizations.
\begin{figure}[h!]
    \centering
    \begin{subfigure}{\linewidth}
        \centering
        \begin{tikzpicture}
        \begin{axis}[
            width=0.8\linewidth,
            xlabel = {$\gamma_t$ (dB)},
            ylabel = {$\overline{R}$},
            xmin = 90, xmax = 110,
            ymin = 5, ymax = 13,
            xtick = {90,95,...,110},
            ytick = {5,7,...,13},
            grid = major,
            legend image post style={xscale=0.9, every mark/.append style={solid}},
            legend cell align = {left},
            legend style={
                at={(0,1)},
                anchor=north west,
                font = \tiny
            }
        ]

        \addplot[
            red,
            no marks,
            line width = 1pt,
            solid
        ]
        table {Fig1_10_1.dat};
        \addlegendentry{$M=1$}
        
        \addplot[
            black,
            no marks,
            line width = 1pt,
            solid
        ]
        table {Fig1_10_2.dat};
        \addlegendentry{$M=2$}
        
        \addplot[
            blue,
            no marks,
            line width = 1pt,
            solid
        ]
        table {Fig1_10_10.dat};
        \addlegendentry{$M=10$}

        \addplot[
            black,
            only marks,
            mark=triangle*,
            mark size=3,
            mark indices={2},
        ]
        table {Fig1_10_2.dat};
        \addlegendentry{Simulation Results}
                
        \addplot[
            black,
            only marks,
            mark=triangle*,
            mark size=3,
            mark indices={0, 1, 2, 3, 5, 8, 11, 15, 25, 44, 73},
        ]
        table {Fig1_10_2.dat};
        
        \addplot[
            red,
            only marks,
            mark=triangle*,
            mark size = 3,
            mark indices={0, 1, 2, 3, 5, 8, 11, 15, 25, 44, 73},
        ]
        table {Fig1_10_1.dat};

        \addplot[
            blue,
            only marks,
            mark=triangle*,
            mark size = 3,
            mark indices={0, 1, 2, 3, 4, 6, 9, 12, 16, 26, 45, 74},
        ]
        table {Fig1_10_10.dat};
        \end{axis}
        \end{tikzpicture}
        \caption{}
    \end{subfigure}
\begin{subfigure}{\linewidth}
        \centering
        \begin{tikzpicture}
        \begin{axis}[
            width=0.8\linewidth,
            xlabel = {$\gamma_t$ (dB)},
            ylabel = {$\overline{R}$},
            xmin = 90, xmax = 110,
            ymin = 3, ymax = 13,
            xtick = {90,95,...,110},
            ytick = {3,5,...,13},
            grid = major,
            legend image post style={xscale=0.9, every mark/.append style={solid}},
            legend cell align = {left},
            legend style={
                at={(0,1)},
                anchor=north west,
                font = \tiny
            }
        ]

        \addplot[
            red,
            no marks,
            line width = 1pt,
            solid
        ]
        table {Fig2_30_1.dat};
        \addlegendentry{$M=1$}
        
        \addplot[
            black,
            no marks,
            line width = 1pt,
            solid
        ]
        table {Fig2_30_2.dat};
        \addlegendentry{$M=2$}
        
        \addplot[
            blue,
            no marks,
            line width = 1pt,
            solid
        ]
        table {Fig2_30_10.dat};
        \addlegendentry{$M=10$}

        \addplot[
            black,
            only marks,
            mark=triangle*,
            mark size=3,
            mark indices={2},
        ]
        table {Fig2_30_2.dat};
        \addlegendentry{Simulation Results}
                
        \addplot[
            black,
            only marks,
            mark=triangle*,
            mark size=3,
            mark indices={0, 1, 2, 3, 5, 8, 11, 15, 25, 44, 73},
        ]
        table {Fig2_30_2.dat};
        
        \addplot[
            red,
            only marks,
            mark=triangle*,
            mark size = 3,
            mark indices={0, 1, 2, 3, 5, 8, 11, 15, 25, 44, 73},
        ]
        table {Fig2_30_1.dat};
        \addplot[
            blue,
            only marks,
            mark=triangle*,
            mark size = 3,
            mark indices={0, 1, 2, 3, 5, 8, 11, 15, 25, 44, 73},
        ]
        table {Fig2_30_10.dat};
        \end{axis}
        \end{tikzpicture}
        \caption{}
    \end{subfigure}
    \caption{Ergodic data rate versus $\gamma_t$ for a two-state PAS for various $M$ values and a) $D_x=10$ m, and b) $D_x=30$ m.}
    \label{Fig2}
\end{figure}
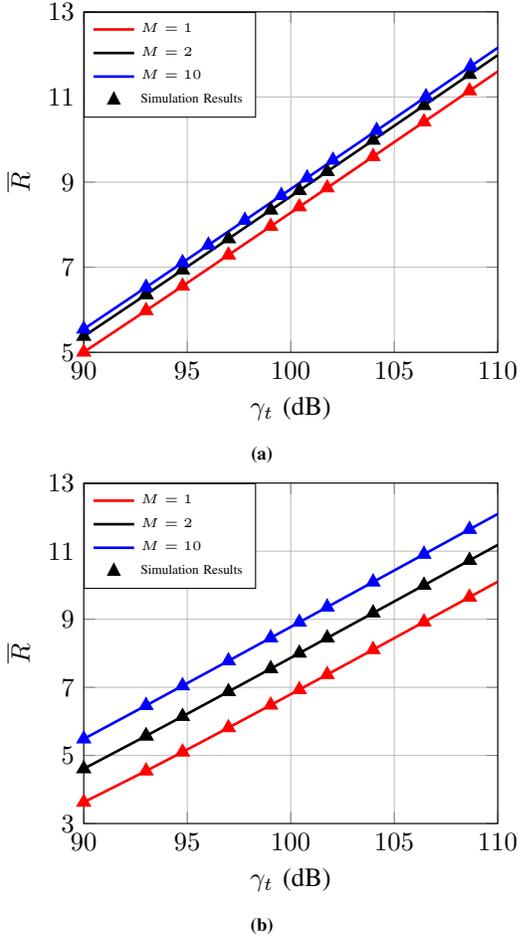

Figs.~\ref{Fig2}a and \ref{Fig2}b illustrate the ergodic data rate of the two-state PAS as a function of $\gamma_t$, for different numbers of PAs, where Fig.~\ref{Fig2}a corresponds to a deployment width of $D_x = 10$~m and Fig.~\ref{Fig2}b to $D_x = 30$~m. Initially, in both cases, the theoretical curves obtained from the derived closed-form expression follow the Monte Carlo simulation results, validating the accuracy of the developed analytical framework across the entire SNR range. Additionally, as shown in Fig.~\ref{Fig2}a, for smaller room dimensions, the achievable data rate exhibits marginal sensitivity to the number of PAs, which shows that once a small number of antennas is employed, the ergodic data rate quickly saturates, and further increasing the number of PAs provides incremental improvements. In contrast, as depicted in Fig.~\ref{Fig2}b, when $D_x$ increases, the separation between the curves becomes more pronounced and the achievable rate improves significantly with $M$, showing that a larger number of PAs becomes beneficial for wider environments where the spatial separation $\delta$ between PAs increases and additional PAs can more effectively compensate for the larger propagation distances. Consequently, Figs.~\ref{Fig2}a and \ref{Fig2}b highlight that the gains from additional PAs become increasingly significant as the deployment area expands.

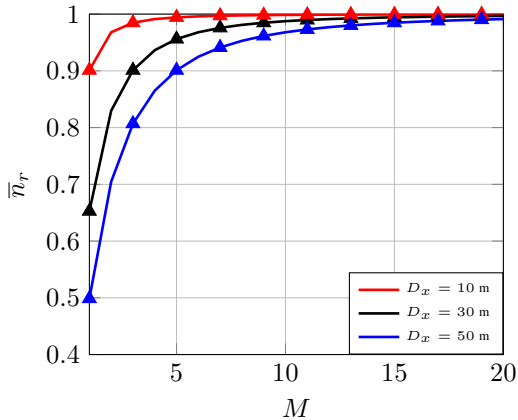
\begin{figure}
    \centering
    \begin{subfigure}{\linewidth}
        \centering
        \begin{tikzpicture}
        \begin{axis}[
            width=0.8\linewidth,
            xlabel = {$M$},
            ylabel = {$\overline{n}_r$},
            xmin = 1, xmax = 20,
            ymin = 0.4, ymax = 1,
            ytick = {0.4,0.5,...,1},
            grid = major,
            legend image post style={xscale=0.9, every mark/.append style={solid}},
            legend cell align = {left},
            legend style={
                at={(1,0)},
                anchor=south east,
                font = \tiny
            }
        ]

        \addplot[
            red,
            no marks,
            line width = 1pt,
            solid
        ]
        table {Fig3_10.dat};
        \addlegendentry{$D_x=10$ m}
        
        \addplot[
            black,
            no marks,
            line width = 1pt,
            solid
        ]
        table {Fig3_30.dat};
        \addlegendentry{$D_x=30$ m}
        
        \addplot[
            blue,
            no marks,
            line width = 1pt,
            solid
        ]
        table {Fig3_50.dat};
        \addlegendentry{$D_x=50$ m}
                
        \addplot[
            black,
            only marks,
            mark=triangle*,
            mark size=3,
            mark repeat=2,
        ]
        table {Fig3_30.dat};
        
        \addplot[
            red,
            only marks,
            mark=triangle*,
            mark size = 3,
            mark repeat=2,
        ]
        table {Fig3_10.dat};

        \addplot[
            blue,
            only marks,
            mark=triangle*,
            mark size = 3,
            mark repeat=2,
        ]
        table {Fig3_50.dat};
        \end{axis}
        \end{tikzpicture}            
    \end{subfigure}
    \caption{PDA versus $M$ for $\gamma_t=90$ dB.}
    \label{Fig3}
\end{figure}

Fig.~\ref{Fig3} depicts the PDE as a function of the number of PAs $M$ for $\gamma_t = 90$~dB. As expected, the PDE increases monotonically with $M$, showing how the discrete PA configuration progressively approaches the ideal continuous case as the number of available PAs grows. In more detail, for compact deployments, such as $D_x = 10$~m, the efficiency rapidly saturates, reaching over $95\%$ of the continuous PAS performance with only two antennas, which reveals that, in small-scale environments, the discretization of the pinching process introduces almost no performance degradation. However, as room dimensions increase, the PDE becomes more sensitive to the number of PAs, and a larger $M$ is required to achieve the same performance level, which arises because increasing $D_x$ leads to a larger spacing $\delta$, which reduces the system’s ability to compensate for the path loss. Nevertheless, even for larger rooms, such as $D_x = 50$~m, the efficiency quickly converges beyond a moderate number of PAs, confirming that near-continuous performance can be obtained with a practical and limited number of PAs, suggesting that continuous PA implementations may not be necessary in practice.
\vspace{-3mm}
\section{Conclusion}
\vspace{-1mm}
In this work, a detailed analytical framework was developed to characterize the performance of two-state PASs as a realization of FAS, through the derivation of an exact closed-form expression for the ergodic rate. In particular, the derived expression accounts for the discrete nature of the PA positions along the software-controllable dielectric waveguide, enabling a direct comparison with the ideal continuous PAS benchmark. Based on this analysis, the performance ratio between the discrete and continuous configurations, namely PDE, was introduced to quantify the performance loss due to spatial discretization. The results reveal that for small deployment widths, the ergodic data rate rapidly saturates with the number of PAs, whereas for larger waveguide lengths, additional PAs become increasingly beneficial in compensating for extended propagation distances. Moreover, near-continuous performance can be achieved with only a finite number of PAs, confirming the practical relevance of two-state PASs. Overall, this work provides theoretical insight and quantitative design guidelines for the efficient deployment of PASs within next-generation FAS.
\vspace{-2mm}
\appendices
\section{Calculation of Integrals $I_i$ and $I_j$}\label{App:A}
Below we provide the calculations for both $I_i$ and $I_j$ integrals:
\subsubsection{Integral $I_i$}
By setting $q = x+\tfrac{\delta^2}{4}$ and applying integration by parts with \(u=\ln(q+y_m^2)\) and \(dv=dy_m\) yields
\begin{equation}
I_i(x)=\frac{\delta}{2}\Big[y_m\ln(q+y_m^2)\Big]_0^{D_y/2}
-\delta\int_{0}^{D_y/2}\frac{y_m^2}{q+y_m^2}\,dy_m,
\end{equation}
which can be rewritten as
\begin{equation}\label{I1_15}
\small
\begin{split}
    I_i(x)= & \frac{\delta}{2}\Big[y_m\ln(q+y_m^2)\Big]_0^{D_y/2}
-\delta\int_{0}^{D_y/2}1\,dy_m  \\& + \delta\int_{0}^{D_y/2}\frac{q}{q+y_m^2}\,dy_m.
\end{split}
\end{equation}
Moreover, by setting $u=\frac{y_m}{\sqrt{q}}$, \eqref{I1_15} can be written as
\begin{equation}\label{I1_16}
\small
\begin{split}
    I_i(x)= \frac{\delta D_y}{4}
    \ln\!\left(\frac{D_y^2}{4}+q\right)
\!-\frac{\delta D_y}{2} 
+ \delta\sqrt{q}\!\int_{0}^{\frac{D_y}{2\sqrt{q}}}\!\frac{1}{1+u^2}\,du.
\end{split}
\end{equation}
Finally, by utilizing \cite[Eq.~(2.01/15)]{GradshteynRyzhik2014}, \eqref{Iix} can be derived, which completes the derivation of $I_i$.
\subsubsection{Integral $I_j$}
By setting $\tan^{-1} \! \left( \frac{\delta}{2 \sqrt{x+ y^2_m}}\right)=u$ and $dv=2 \sqrt{x+ y^2_m}\,dy_m$ and applying integration by parts, \eqref{Ij_13} can be written as
\begin{equation}\label{Ij_17}
\small
\begin{aligned}
&I_j(x)= \!\int_{0}^{\frac{D_y}{2}}
\frac{2\delta\,y_m
\Big(y_m\sqrt{x+y_m^2}
+ x\ln\!\big(y_m+\sqrt{x+y_m^2}\big)\!\Big)}
{\sqrt{x+y_m^2}\,\big(4(x+y_m^2)+\delta^2\big)}\,dy_m \\
& \!+\!\Bigg[\!
\tan^{-1}\!\left(\!\tfrac{\delta}{2\sqrt{x+y_m^2}}\!\right)
\!\Big(y_m\!\sqrt{x+y_m^2}
\!+\! x\ln\!\Big(y_m+ \sqrt{x+y_m^2}\Big)\!\Big)  \!\Bigg]_{0}^{\!\frac{D_y}{2}}\!,
\end{aligned}
\end{equation}
which, after some algebraic manipulations, can be expressed as
\begin{equation}\label{Ij_18}
\small
\begin{aligned}
&I_j(x)\!= \!\tan^{-1}\!\left(\!\frac{\delta}{2\sqrt{x+\tfrac{D_y^2}{4}}}\!\right)
\!\Bigg(\!\frac{D_y}{2}\sqrt{x+\tfrac{D_y^2}{4}}
+ \\&x\ln\!\Big(\tfrac{D_y}{2}+\sqrt{x+\tfrac{D_y^2}{4}}\Big)\!\Bigg) 
- x\ln(\sqrt{x})\,\tan^{-1}\!\left(\frac{\delta}{2\sqrt{x}}\right)\!+ \\[2pt]
& \!\underbrace{\displaystyle\int\limits_{0}^{\frac{D_y}{2}}\!
\frac{\delta\,y_m^2}{2(x+y_m^2+\frac{\delta^2}{4})}dy_m}_{\textstyle J_1}
+
\underbrace{\displaystyle\int\limits_{0}^{\frac{D_y}{2}}
\frac{\delta\,x\,y_m\,\ln\!\big(y_m+\sqrt{x+y_m^2}\big)}
{2\sqrt{x+y_m^2}\,\big(x+y_m^2+\frac{\delta^2}{4}\big)}dy_m}_{\textstyle J_2}\!.
\end{aligned}
\end{equation}
The evaluation of $I_j(x)$ then proceeds through the calculation of $J_1$ and $J_2$, each of which can be expressed in closed form.

\underline{\textit{i) Calculation of $J_1$:}} Initially, to calculate $J_1$ we can reformulate it as follows
\begin{equation}
\small
    J_1=\frac{\delta}{2}\int_{0}^{\frac{D_y}{2}}\!1-
\frac{x+ \frac{\delta^2}{4}}{x+y_m^2+\frac{\delta^2}{4}}dy_m,
\end{equation}
Moreover, by setting $q = x+\tfrac{\delta^2}{4}$ and $u = \tfrac{y_m}{\sqrt{q}}$, after some algebraic manipulations, we obtain
\begin{equation}
\small
    J_1=\frac{\delta}{2}\!\left[\frac{D_y}{2}
    - \sqrt{q}\!\int_{0}^{\frac{D_y}{2\sqrt{q}}}\!\frac{1}{1+u^2}\,du\right].
\end{equation}
Finally, by using \cite[(2.01/15)]{GradshteynRyzhik2014}, yields
\begin{equation}\label{J1_final}
\small
    J_1 = \frac{\delta D_y}{4}
    - \frac{\delta}{2}\sqrt{x+\frac{\delta^2}{4}}\,
    \tan^{-1}\!\left(\frac{D_y}{2\sqrt{x+\frac{\delta^2}{4}}}\right),
\end{equation}
which completes the calculation of $J_1$.

\underline{\textit{ii) Calculation of $J_2$:}} 
By setting $y_m=\!\sqrt{x}\sinh(u)$, then $J_2$ can be expressed as
\begin{equation}\label{J2_22}
\small
J_2= \displaystyle\int\limits_{0}^{\operatorname{asinh}\!\left(\frac{D_y}{2\sqrt{x}}\right)} 
    \frac{\delta x\sqrt{x}\sinh(u)\!\ln\!\Big(\!\sqrt{x}\sinh(u)\!+\!\sqrt{x}\cosh(u)\!\Big)}
    {2\big(x\cosh^2(u)+\frac{\delta^2}{4}\big)}du.
\end{equation}
Additionally, by utilizing the hyperbolic identity $\sinh(u)+\cosh(u)=e^{u}$, after some algebraic manipulations, \eqref{J2_22} can be rewritten as
\begin{equation}
\small
J_2= \frac{\delta \sqrt{x}}{2}\displaystyle\int\limits_{0}^{\operatorname{asinh}\left(\frac{D_y}{2\sqrt{x}}\right)}
\frac{\sinh (u)\big(\ln(\sqrt{x})+u\big)}
{\cosh^2(u)+\frac{\delta^2}{4x}}\,du.
\end{equation}
By applying integration by parts with 
$s=\ln(\sqrt{x})+u$ and 
$dw=\frac{\delta\sqrt{x}}{2}\frac{\sinh u}{\cosh^2 u+\frac{\delta^2}{4x}}\,du$, 
we obtain
\begin{equation}\label{J2_24}
\small
\begin{split}
J_2= &\left[x\Big(\ln(\sqrt{x})+u\Big)\tan^{-1}\!\Big(\tfrac{2\sqrt{x}}{\delta}\cosh(u)\Big)
\right]_{0}^{\operatorname{asinh}\!\left(\frac{D_y}{2\sqrt{x}}\right)} \\&
- x\!\int_{0}^{\operatorname{asinh}\!\left(\frac{D_y}{2\sqrt{x}}\right)}
\tan^{-1}\!\Big(\tfrac{2\sqrt{x}}{\delta}\cosh(u)\Big)\,du.
\end{split}
\end{equation}
By utilizing the identity $\cosh(\operatorname{asinh} z)=\sqrt{1+z^2}$, then \eqref{J2_24} simplifies to
\begin{equation}\label{J2_25}
\small
\begin{aligned}
J_2 & = x\Bigg(\!\Big(\!\ln(\sqrt{x})+\operatorname{asinh}\!\left(\tfrac{D_y}{2\sqrt{x}}\right)\!\Big)\!
\tan^{-1}\!\left(\!\tfrac{\sqrt{4x+D_y^2}}{\delta}\!\right) \!-\!\ln(\sqrt{x}) \\
& \times\tan^{-1}\!\Big(\tfrac{2\sqrt{x}}{\delta}\Big)
-\int_{0}^{\operatorname{asinh}\!\left(\frac{D_y}{2\sqrt{x}}\right)}
\tan^{-1}\!\Big(\tfrac{2\sqrt{x}}{\delta}\cosh(u)\Big)du \Bigg).
\end{aligned}
\end{equation}
By using the expansion $\tan^{-1}(z)=\tfrac{j}{2}\big[\ln(1-jz)-\ln(1+jz)\big]$, then \eqref{J2_25} can be reformulated as
\begin{equation}\label{J2_26}
\small
\begin{aligned}
J_2 &= x\Bigg(\!\Big(\!\ln(\sqrt{x})+\operatorname{asinh}\!\left(\tfrac{D_y}{2\sqrt{x}}\right)\!\Big)\!
\tan^{-1}\!\left(\!\tfrac{\sqrt{4x+D_y^2}}{\delta}\!\right) \!-\!\ln(\sqrt{x}) \\
& \times\tan^{-1}\!\Big(\tfrac{2\sqrt{x}}{\delta}\Big)
- \frac{j}{2}\!\displaystyle\int\limits_{0}^{\operatorname{asinh}\!\left(\tfrac{D_y}{2\sqrt{x}}\right)}\!
\ln\!\Bigg(\frac{1-j\tfrac{2\sqrt{x}}{\delta}\cosh(u)}{1+j\tfrac{2\sqrt{x}}{\delta}\cosh(u)}\Bigg)du
\Bigg).
\end{aligned}
\end{equation}
Additionally, by setting $\xi=e^{u}$ and after some algebraic manipulations, \eqref{J2_26} becomes
\begin{equation}\label{J2_27}
\small
\begin{aligned}
J_2 & = x\Bigg(\!Big(\!\ln(\sqrt{x})+\operatorname{asinh}\!\left(\tfrac{D_y}{2\sqrt{x}}\right)\!\Big)\!
\tan^{-1}\!\left(\!\tfrac{\sqrt{4x+D_y^2}}{\delta}\!\right) \!-\!\ln(\sqrt{x}) \\
& \times\tan^{-1}\!\Big(\tfrac{2\sqrt{x}}{\delta}\Big)
- \frac{j}{2}\int\limits_{1}^{e^{A}}
\ln\!\Bigg(\frac{2\xi - j\,\tfrac{2\sqrt{x}}{\delta}\,(\xi^2+1)}
     {2\xi + j\,\tfrac{2\sqrt{x}}{\delta}\,(\xi^2+1)}\Bigg)\ \frac{d\xi}{\xi}\Bigg),
\end{aligned}
\end{equation}
with $A={\mathrm{asinh}}\left(\tfrac{D_y}{2\sqrt{x}}\right)$, and by factorizing the quadratic terms in the numerator and denominator of the logarithm in \eqref{J2_27}, the logarithm can be expressed as four elementary logarithms, resulting in
\begin{equation}\label{J2_28}
\small
\begin{aligned}
&J_2= x\Bigg(\!\Big(\!\ln(\sqrt{x})+\operatorname{asinh}\!\left(\tfrac{D_y}{2\sqrt{x}}\right)\!\Big)\!
\tan^{-1}\!\left(\!\tfrac{\sqrt{4x+D_y^2}}{\delta}\!\right) \!-\!\ln(\sqrt{x}) \\
&\!\times\tan^{-1}\!\Big(\tfrac{2\sqrt{x}}{\delta}\Big)
- \frac{j}{2}\int\limits_{1}^{e^A}\!\
\ln\!\Big(\xi + j\,\tfrac{\delta}{2\sqrt{x}}\big(1+\sqrt{1+\tfrac{4x}{\delta^2}}\big)\Big) \\
&\!+\!\ln\!\Big(\xi - j\,\tfrac{\delta}{2\sqrt{x}}\big(\sqrt{1+\tfrac{4x}{\delta^2}}-1\big)\Big)\!-\!\ln\!\Big(\xi - j\,\tfrac{\delta}{2\sqrt{x}}\big(1-\sqrt{1+\tfrac{4x}{\delta^2}}\big)\Big) \\
&\!-\ln\!\Big(\xi - j\,\tfrac{\delta}{2\sqrt{x}}\big(1+\sqrt{1+\tfrac{4x}{\delta^2}}\big)\Big)
\Big]\frac{d\xi}{\xi}\Bigg).
\end{aligned}
\end{equation}
Moreover, by applying the identity $\ln(\xi - a)=\ln\xi+\ln\!\big(1-\tfrac{a}{\xi}\big)$, \eqref{J2_28} can be rewritten as
\begin{equation}\label{J2_29}
\small
\begin{aligned}
&J_2= x\Bigg(\!\Big(\!\ln(\sqrt{x})+\operatorname{asinh}\!\left(\tfrac{D_y}{2\sqrt{x}}\right)\!\Big)\!
\tan^{-1}\!\left(\!\tfrac{\sqrt{4x+D_y^2}}{\delta}\!\right) \!-\!\ln(\sqrt{x}) \\
&\times\tan^{-1}\!\Big(\tfrac{2\sqrt{x}}{\delta}\Big)
- \frac{j}{2}\!\int\limits_{1}^{e^A}\!
\Big[
\ln\!\Big(1-\tfrac{-\,j\,\tfrac{\delta}{2\sqrt{x}}\big(1+\sqrt{1+\tfrac{4x}{\delta^2}}\big)}{\xi}\Big) \\
&+\ln\!\Big(1-\tfrac{+\,j\,\tfrac{\delta}{2\sqrt{x}}\big(\sqrt{1+\tfrac{4x}{\delta^2}}-1\big)}{\xi}\Big)
-\ln\!\Big(1-\tfrac{+\,j\,\tfrac{\delta}{2\sqrt{x}}\big(1-\sqrt{1+\tfrac{4x}{\delta^2}}\big)}{\xi}\Big) \\
&\!-\ln\!\Big(1-\tfrac{+\,j\,\tfrac{\delta}{2\sqrt{x}}\big(1+\sqrt{1+\tfrac{4x}{\delta^2}}\big)}{\xi}\Big)
\Big]\frac{d\xi}{\xi}\Bigg).
\end{aligned}
\end{equation}
By taking into account that $\int \frac{\ln\!\big(1-\tfrac{a}{\xi}\big)}{\xi}\,d\xi=\operatorname{Li}_2\!\Big(\tfrac{a}{\xi}\Big)$, we obtain
\begin{equation}\label{J2_30}
\small
\begin{aligned}
&J_2= x\Bigg(\!\Big(\!\ln(\sqrt{x})+\operatorname{asinh}\!\left(\tfrac{D_y}{2\sqrt{x}}\right)\!\Big)\!
\tan^{-1}\!\left(\!\tfrac{\sqrt{4x+D_y^2}}{\delta}\!\right) \!-\!\ln(\sqrt{x}) \\
&\times\tan^{-1}\!\Big(\tfrac{2\sqrt{x}}{\delta}\Big)
- \frac{j}{2}\Bigg[
\operatorname{Li}_2\!\Big(\tfrac{-\,j\,\tfrac{\delta}{2\sqrt{x}}\big(1+\sqrt{1+\tfrac{4x}{\delta^2}}\big)}{\xi}\Big)
\\
&+\operatorname{Li}_2\!\Big(\tfrac{+\,j\,\tfrac{\delta}{2\sqrt{x}}\big(\sqrt{1+\tfrac{4x}{\delta^2}}-1\big)}{\xi}\Big)-\operatorname{Li}_2\!\Big(\tfrac{+\,j\,\tfrac{\delta}{2\sqrt{x}}\big(1-\sqrt{1+\tfrac{4x}{\delta^2}}\big)}{\xi}\Big) \\
&-\operatorname{Li}_2\!\Big(\tfrac{+\,j\,\tfrac{\delta}{2\sqrt{x}}\big(1+\sqrt{1+\tfrac{4x}{\delta^2}}\big)}{\xi}\Big)
\Bigg]_{1}^{e^{A}}\Bigg),
\end{aligned}
\end{equation}
and by using the identities $\operatorname{Li}_2\!\Big(\tfrac{1}{z}\Big)
= -\operatorname{Li}_2(z)-\frac{\pi^2}{6}-\frac{1}{2}\,\ln^2(-z)$, and $\ln(-z)=\ln z + j\pi$, after some algebraic manipulations we obtain
\begin{equation}\label{J2_31}
\small
\begin{aligned}
&J_2= x\Bigg(\!\Big(\!\ln(\sqrt{x})+\operatorname{asinh}\!\left(\tfrac{D_y}{2\sqrt{x}}\right)\!\Big)\!
\tan^{-1}\!\left(\!\tfrac{\sqrt{4x+D_y^2}}{\delta}\!\right) \!-\!\ln(\sqrt{x}) \\
&\!\times\tan^{-1}\!\Big(\tfrac{2\sqrt{x}}{\delta}\Big)
-\frac{\pi}{2}\operatorname{asinh}\!\left(\tfrac{D_y}{2\sqrt{x}}\right) \\
&\!-\frac{j}{2}\Bigg[
\operatorname{Li}_2\!\Big(\tfrac{-j\delta}{2\sqrt{x}}\!\left(1+\sqrt{1+\tfrac{4x}{\delta^2}}\right)e^{-\!A}\Big)
\\&\!-\operatorname{Li}_2\!\Big(\tfrac{-j\delta}{2\sqrt{x}}\!\left(1+\sqrt{1+\tfrac{4x}{\delta^2}}\right)e^{A}\Big)
-\operatorname{Li}_2\!\Big(\tfrac{j\delta}{2\sqrt{x}}\!\left(1+\sqrt{1+\tfrac{4x}{\delta^2}}\right)e^{-\!A}\Big)
\\
&\!+\operatorname{Li}_2\!\Big(\tfrac{j\delta}{2\sqrt{x}}\!\left(1+\sqrt{1+\tfrac{4x}{\delta^2}}\right)e^{A}\Big)
\Bigg]\Bigg).
\end{aligned}
\end{equation}
Finally, by utilizing the definition of the arctangent integral function $\mathrm{Ti}_2(z)$, we obtain \eqref{J2_32}, which concludes the derivation of $J_2$.
\begin{figure*}[t]
\centering
\begin{equation}\label{J2_32}
\small
\begin{aligned}
&J_2= x\!\Bigg[\!\Big(\!\ln(\sqrt{x})+\operatorname{asinh}\!\left(\tfrac{D_y}{2\sqrt{x}}\right)\!\Big)\!
\tan^{-1}\!\left(\!\tfrac{\sqrt{4x+D_y^2}}{\delta}\!\right) \!-\!\ln(\sqrt{x})\tan^{-1}\!\Big(\tfrac{2\sqrt{x}}{\delta}\Big)
-\tfrac{\pi}{2}\operatorname{\mathrm{asinh}}\!\left(\tfrac{D_y}{2\sqrt{x}}\right) 
-\mathrm{Ti}_2\Big(
e^{\operatorname{\mathrm{asinh}}\!\left(\tfrac{D_y}{2\sqrt{x}}\right)}
\tfrac{\delta}{2\sqrt{x}}
\!\left(\sqrt{1+\tfrac{4x}{\delta^2}}-1\right)\!\Big) \\&
+\mathrm{Ti}_2\Big(
\tfrac{\delta}{2\sqrt{x}}
\!\left(\sqrt{1+\tfrac{4x}{\delta^2}}-1\right)\!\Big)
-\mathrm{Ti}_2\Big(
-\,e^{\operatorname{asinh}\!\left(\tfrac{D_y}{2\sqrt{x}}\right)}
\tfrac{\delta}{2\sqrt{x}}
\!\left(\sqrt{1+\tfrac{4x}{\delta^2}}+1\right)\!\Big)+\mathrm{Ti}_2\Big(
-\,\tfrac{\delta}{2\sqrt{x}}
\!\left(\sqrt{1+\tfrac{4x}{\delta^2}}+1\right)\!\Big)
\Bigg],
\end{aligned}
\end{equation}
\vspace{2mm}
\hrule
\end{figure*}

\section*{Acknowledgment}
This work has been funded by the European Union’s Horizon 2020 research and innovation programs under grant agreement No 101139194 6G Trans-Continental Edge Learning.
\bibliographystyle{IEEEtran}
\bibliography{Bibliography}

@book{GradshteynRyzhik2014,
  author    = {I. S. Gradshteyn and I. M. Ryzhik},
  title     = {Table of Integrals, Series, and Products},
  publisher = {Academic Press},
  address   = {New York, NY, USA},
  year      = {2014}
}

@ARTICLE{Ding2024TCOM,
  author={Ding, Zhiguo and Schober, Robert and Vincent Poor, H.},
  journal={IEEE Trans. Commun.}, 
  title={Flexible-Antenna Systems: {A} Pinching-Antenna Perspective}, 
  year={2025},
  volume={},
  number={},
  pages={1--1},
  doi={10.1109/TCOMM.2025.3555866}}

@ARTICLE{6GNetwork,
  author={Jiang, Hao and Mukherjee, Mithun and Zhou, Jie and Lloret, Jaime},
  journal={IEEE Netw.}, 
  title={Channel Modeling and Characteristics for {6G} Wireless Communications}, 
  year={2021},
  volume={35},
  number={1},
  pages={296-303},
  doi={10.1109/MNET.011.2000348}
}

@ARTICLE{DOCOMO,
  author = {Fukuda, A. and Yamamoto, H. and Okazaki, H. and Suzuki, Y. and Kawai, K.},
  title = {Pinching Antenna-Using a Dielectric Waveguide as an Antenna},
  journal = {NTT DOCOMO Tech. J.},
  volume = {23},
  number = {3},
  pages = {5-12},
  month = {Jan.},
  year = {2022}
}

@ARTICLE{TegosPinching,
  author={Tegos, Sotiris A. and Diamantoulakis, Panagiotis D. and Ding, Zhiguo and Karagiannidis, George K.},
  journal={IEEE Wireless Commun. Lett.}, 
  title={Minimum Data Rate Maximization for Uplink Pinching-Antenna Systems}, 
  year={2025},
  volume={},
  number={},
  pages={1-1},
  doi={10.1109/LWC.2025.3547956}}

@ARTICLE{PASS,
  author={Ouyang, Chongjun and Wang, Zhaolin and Liu, Yuanwei and Ding, Zhiguo},
  journal={IEEE Commun. Lett.}, 
  title={Array Gain for Pinching-Antenna Systems {(PASS)}}, 
  year={2025},
  volume={29},
  number={6},
  pages={1471--1475},
  doi={10.1109/LCOMM.2025.3566299}}

@ARTICLE{Modeling2025,
  author={Wang, Zhaolin and Ouyang, Chongjun and Mu, Xidong and Liu, Yuanwei and Ding, Zhiguo},
  journal={IEEE Trans. Commun.}, 
  title={Modeling and Beamforming Optimization for Pinching-Antenna Systems}, 
  year={2025},
  volume={},
  number={},
  pages={1--1},
  doi={10.1109/TCOMM.2025.3621049}}

@ARTICLE{TyrovolasPASS2025,
  author={Tyrovolas, Dimitrios and Tegos, Sotiris A. and Diamantoulakis, Panagiotis D. and Ioannidis, Sotiris and Liaskos, Christos K. and Karagiannidis, George K.},
  journal={IEEE Trans. Cogn. Commun. Netw.}, 
  title={Performance Analysis of Pinching-Antenna Systems}, 
  year={2025},
  volume={},
  number={},
  pages={1--1},
  doi={10.1109/TCCN.2025.3564470}}

@ARTICLE{VasilisPASS,
  author={Papanikolaou, Vasilis K. and Zhou, Gui and Kaziu, Brikena and Khalili, Ata and Diamantoulakis, Panagiotis D. and Karagiannidis, George K. and Schober, Robert},
  journal={IEEE Wireless Commun. Lett.}, 
  title={Resolving the Double Near-Far Problem via Wireless Powered Pinching-Antenna Networks}, 
  year={2025},
  volume={},
  number={},
  pages={1--1},
  doi={10.1109/LWC.2025.3592881}}

@misc{BozanisPASS,
  title={Cram{\'e}r-Rao Bounds for Integrated Sensing and Communications in Pinching-Antenna Systems}, 
  author={Dimitrios Bozanis and Vasilis K. Papanikolaou and Sotiris A. Tegos and George K. Karagiannidis},
  year={2025},
  eprint={2505.01333},
  archivePrefix={arXiv},
  primaryClass={eess.SP},
  url={https://arxiv.org/abs/2505.01333},
}

@ARTICLE{ThrassosPASS,
  author={Oikonomou, Thrassos K. and Tegos, Sotiris A. and Diamantoulakis, Panagiotis D. and Liu, Yuanwei and Karagiannidis, George K.},
  journal={IEEE Commun. Lett.}, 
  title={{OFDMA} for Pinching-Antenna Systems}, 
  year={2025},
  volume={},
  number={},
  pages={1--1},
  doi={10.1109/LCOMM.2025.3614552}}

@misc{APOSTOLOSRSMA,
      title={Uplink {RSMA} for Pinching-Antenna Systems}, 
      author={Apostolos A. Tegos and Yue Xiao and Sotiris A. Tegos and George K. Karagiannidis and Panagiotis D. Diamantoulakis},
      year={2025},
      eprint={2509.10076},
      archivePrefix={arXiv},
      primaryClass={eess.SP},
      url={https://arxiv.org/abs/2509.10076}, 
}

@misc{PIGIPASS,
      title={Secrecy Rate Maximization with Artificial Noise for Pinching-Antenna Systems}, 
      author={Pigi P. Papanikolaou and Dimitrios Bozanis and Sotiris A. Tegos and Panagiotis D. Diamantoulakis and George K. Karagiannidis},
      year={2025},
      eprint={2504.10656},
      archivePrefix={arXiv},
      primaryClass={eess.SP},
      url={https://arxiv.org/abs/2504.10656}, 
}

@INPROCEEDINGS{Assimonis,
  author={Assimonis, Stylianos D.},
  booktitle={2023 IEEE SENSORS}, 
  title={How Challenging is it to Design a Practical Superdirective Antenna Array?}, 
  year={2023},
  volume={},
  number={},
  pages={1-4},
  doi={10.1109/SENSORS56945.2023.10325299}}

@ARTICLE{Kaikit2025,
  author={Chen, Jung-Chieh and Wu, Po-Ching and Wong, Kai-Kit},
  journal={IEEE Open J. Commun. Soc.}, 
  title={Dynamic Placement of Pinching Antennas for Multicast {MU-MISO} Downlinks}, 
  year={2025},
  volume={6},
  number={},
  pages={5611-5625},
  doi={10.1109/OJCOMS.2025.3582895}}

@ARTICLE{FAS2021,
  author={Wong, Kai-Kit and Shojaeifard, Arman and Tong, Kin-Fai and Zhang, Yangyang},
  journal={IEEE Trans. Wireless Commun.}, 
  title={Fluid Antenna Systems}, 
  year={2021},
  volume={20},
  number={3},
  pages={1950-1962},
  doi={10.1109/TWC.2020.3037595}}
\end{document}